# Revolving scheme for solving a cascade of Abel equations in dynamics of planar satellite rotation

**Sergey V. Ershkov**

Sternberg Astronomical Institute,

M.V. Lomonosov's Moscow State University,

13 Universitetskij prospect, Moscow 119992, Russia

e-mail: sergej-ershkov@yandex.ru

The main objective for this research was the analytical exploration of the dynamics of planar satellite rotation during the motion of an elliptical orbit around a planet.

First, we revisit the results of J. Wisdom *et al.* (1984), in which, by the elegant change of variables (considering the true anomaly *f* as the *independent* variable), the governing equation of satellite rotation takes the form of an *Abel* ODE of the second kind, a sort of generalization of the *Riccati* ODE. We note that due to the special character of solutions of a *Riccati*-type ODE, there exists the possibility of sudden *jumping* in the magnitude of the solution at some moment of time.

In the physical sense, this jumping of the *Riccati*-type solutions of the governing ODE could be associated with the effect of sudden acceleration/deceleration in the satellite rotation around the chosen principle axis at a definite moment of parametric time. This means that there exists not only a chaotic satellite rotation regime (as per the results of J. Wisdom *et al.* (1984)), but a kind of *gradient catastrophe* (Arnold 1992) could occur during the satellite rotation process. We especially note that if a gradient catastrophe *could occur*, this does not mean that it must occur: such a possibility depends on the initial conditions.

In addition, we obtained asymptotical solutions that manifest a *quasi-periodic* character even with the strong simplifying assumptions $e \to 0$, $p = 1$, which reduce the governing equation of J. Wisdom *et al.* (1984) to a kind of Beletskii's equation.

# 1. Introduction of the governing equation

The planar oscillations and rotations of a satellite around its center of mass, while moving along an elliptical orbit with eccentricity $e$, is described by Beletskii's equation [1-5]:

$$(1+e\cos f)\frac{d^2\delta}{df^2} - 2e\sin f\frac{d\delta}{df} + \omega_0^2\sin\delta = 4e\sin f \qquad (1)$$

where δ is the *doubled* angle between the radius vector of the center of mass and one of the axes of inertia of the satellite, $\omega_o^2 = 3(B - A)/C$ is the inertial parameter of the satellite, $A$ and $B$ are the moments of inertia of the satellite with respect to the axes of inertia lying in the orbit plane, and $C$ is the moment of inertia with respect to the axis perpendicular to the orbit plane. The true anomaly $f$ is *the independent* variable, which is the angular distance of the radius vector from the pericenter of the orbit.

With reference to the results of [6], consider a satellite whose spin axis is normal to its orbital plane. The satellite is assumed to be a *triaxial* ellipsoid with principal moments of inertia $A < B < C$, and $C$ is the moment about the spin axis. The orbit is assumed to be a fixed ellipse with the semi-major axis $a$, eccentricity $e$, true anomaly $f$, instantaneous radius $r$, and longitude of *periapse* $\varpi$, which is taken as the origin of the longitudes. The orientation of the satellite's long axis is specified by $\vartheta$, thus ($\vartheta - f$) measures the orientation of the satellite's long axis relative to the planet–satellite center line. This notation is the same as that used in [6-7]. Without external tidal torques, the equation of motion for $\vartheta$ is as follows:

$$\frac{d^2\vartheta}{dt^2} + \frac{\omega_0^2}{2r^3}\sin(2(\vartheta - f)) = 0 \qquad (2)$$

where $\omega_o^2 = 3(B - A)/C$ and the units have been chosen so that the orbital period of the satellite is $2\pi$ and its semi-major axis is equal to one. Thus, the dimensionless time $t$ is equal to the mean longitude.

If we define $\psi_p = (\vartheta - p \cdot f)$ (where $p$ is the ratio of frequencies in the spin-orbit resonance) and we rewrite Eq. (2) with the true anomaly $f$ as the *independent* variable, the equation of motion for $\psi_p$ becomes the following (*for the full derivation, see the Appendix; only the resulting formulae are given in the main text below*):

$$(1+e\cos f) \cdot \frac{d^2\psi_p}{df^2} - 2e\sin f \cdot \left(p + \frac{d\psi_p}{df}\right) + \frac{\omega_0^2}{2} \cdot \sin\left(2(\psi_p + (p-1)\cdot f)\right) = 0 \quad (3)$$

As a first approximation, we could transform Eq. (3) by changing the variables $(d\psi_p/df) = \Psi(\psi_p)$ as follows (note that $(d^2\psi_p/df^2) = (d\Psi/d\psi_p)\cdot\Psi(\psi_p)$):

$$\left(D \cdot \Psi\right) \cdot \left(\frac{d\Psi}{d\psi_p}\right) = E \cdot \Psi + F(\psi_p) \quad (4)$$

$$D = (1+e\cos f), \quad E = 2e\sin f, \quad F(\psi_p) = 2e\sin f \cdot p - \frac{\omega_0^2}{2} \cdot \sin\left(2(\psi_p + (p-1)\cdot f)\right)$$

Recall that the true anomaly $f$ is supposed to be an *independent* variable, also independent of $\psi_p$. So, the true anomaly $f$ could be considered to be *a slow variable* parameter in Eq. (4), which is known to periodically change the elliptical motion of the satellite on its orbit around the planet (or *quasi*-periodically).

The above Eq. (4) is known as the *Abel* ODE of the second kind (for research with Abel ODE, see [8]), which is a kind of generalization of the *Riccati* ODE (for research with Riccati ODE, see [9]). This type of equation has no general analytical solution [10]. We also note the existence of a modern method for obtaining the solution of *Riccati* equations with good approximation [11].

## 2. Revolving scheme for solving cascades of Abel equations for satellite rotation

First, we clarify the essential assumption made in the transformation of Eq. (3) to Eq. (4), namely, that the true anomaly $f$ could be considered to be *a slow variable* parameter in Eq. (4), which is known to periodically change the elliptical motion of the satellite on its orbit around the planet (or *quasi*-periodically).

We neglect the tidal torque in the derivation of Eqs. (1) and (2), which can be estimated as shown below [7], [12]:

$$T = -\frac{3}{2}\frac{GM^2}{C}\frac{R^5}{r^6}\frac{k_2}{Q} \qquad (5)$$

where $G$ is the gravitational constant, $M$ is the mass of the planet, $C$ is the *maximal* principal moment of inertia of the satellite, $R$ is the mean volumetric radius of the satellite, $k_2$ is the *Love* number (which describes response of the gravitational potential of the distorted body in regard to influence of tides), $Q$ is the quality factor (which is inversely proportional to the amount of energy dissipated essentially as heat by tidal friction). In addition, we recall that the instantaneous radius $r$ for the Keplerian satellite motion is as follows:

$$r = a \cdot \left(\frac{1-e^2}{1+e\cdot\cos f}\right) \qquad (6)$$

For a *Maclaurin* spheroid satellite, $C = 0.4\ m\cdot R^2$ (where $m$ is the mass of the satellite), we obtain the following:

$$T = -\frac{15}{4}\frac{GM^2}{m}\frac{R^3}{r^6}\frac{k_2}{Q} \qquad (7)$$

So, using Eqs. (5)–(7), Eq. (2) should be presented in its complete form [7] (*we assume GM = 1 below for the sake of computational simplicity*):

$$\frac{d^2\vartheta}{dt^2} + \frac{\omega_0^2}{2}\frac{GM}{r^3}\sin(2(\vartheta - f)) = T \qquad (8)$$

In general, by averaging the tidal torque, Eq. (7), for the satellite period of the full rotation on orbit around the planet, [13] obtained the following for Eq. (8) (see [14]):

$$(1+e\cos f)\frac{d^2\delta}{df^2} + \left(\alpha \cdot (1+e\cos f)^5 - 2e\sin f\right)\frac{d\delta}{df} + \omega_0^2 \cdot \sin\delta = 4e\sin f, \quad (9)$$

$$\alpha = \frac{15}{4}\frac{M}{m}\left(\frac{R}{a}\right)^3 \frac{k_2}{Q}(1-e^2)^{-\frac{9}{2}}$$

where $\delta$ is the *doubled* angle between the radius vector of the center of mass and the appropriate axis of inertia of the satellite.

Now, we could estimate from Eq. (9) the changing $\Delta f$ of the true anomaly $f$ during the motion of the satellite on its orbit, for which $f$ could be considered to be *a constant* parameter when resolving Eq. (4). In such a short period $\Delta f$, the change in the term ($2e \cdot \sin f$) in Eq. (9) should be less than the change in the previously neglected tidal torque during the derivation of Eqs. (1) and (2):

$$\alpha \cdot (1+e\cos f)^5 \cong 2e \cdot \Delta f, \quad \Rightarrow \quad \Delta f \leq \frac{\alpha \cdot (1-e)^5}{2e}, \quad (10)$$

$$\alpha = \frac{15}{4}\frac{M}{m}\left(\frac{R}{a}\right)^3 \frac{k_2}{Q}(1-e^2)^{-\frac{9}{2}}$$

or in case of $e \to 0$,

$$\Delta f \cong \frac{\alpha}{2e}, \quad (11)$$

$$\alpha = \frac{15}{4}\frac{M}{m}\left(\frac{R}{a}\right)^3 \frac{k_2}{Q}$$

where $k_2 \cong 0.03$, $Q \cong 100$ [14] (*see the proper estimations in the Appendix*).

Thus, we obtain the revolving scheme for solving a cascade of Abel equations for calculating satellite rotation during the motion of the satellite on its orbit: the variable parameter of the true anomaly $f$ should be divided by $n$ steps ($n = f/\Delta f$), with the size of each assumed to be $\Delta f$ (10)–(11); during the motion of the satellite on its orbit within each interval Δ*f*, the absolute meaning of the true anomaly *f* could be considered to be *a constant* parameter when resolving Eq. (4). In addition, at each calculation step, the final solution values of the previous step should be considered to be the *initial conditions* for the next calculation step.

### 3. Asymptotical solution of the governing ODE (case *e* → *0*)

We note that in the *p* = 1 case, Eq. (3) could be transformed into a kind of Beletskii's Eq. (1) by changing the variables $\psi_p \to \delta$, where δ is the *doubled* angle between the radius vector of the center of mass and the appropriate axis of inertia of the satellite. So, we should multiply both Eqs. (2) and (3) by two to transform it into a Beletskii's equation, along with the corresponding inertial parameter of the satellite, $\omega_0^2$ (as well as the corresponding moments of inertia of the satellite with respect to the axis of inertia lying in the orbit plane).

Let us transform Eq. (4) from an *Abel* ODE of the second kind to an *Abel* ODE of the first kind [10], by changing the variables Ψ = 1/Ω, as follows:

$$\left( D \cdot \frac{1}{\Omega} \right) \cdot \left( -\frac{1}{\Omega^2} \cdot \frac{d\Omega}{d\psi_p} \right) = E \cdot \frac{1}{\Omega} + F(\psi_p) \quad \Rightarrow$$

$$\frac{d\Omega}{d\psi_p} = \left( -\frac{F(\psi_p)}{D} \right) \cdot \Omega^3 - \frac{E}{D} \cdot \Omega^2 \tag{12}$$

$$D = (1 + e\cos f), \quad E = 2e\sin f, \quad F(\psi_p) = 2e\sin f \cdot p - \frac{\omega_0^2}{2} \cdot \sin\left( 2(\psi_p + (p-1)\cdot f) \right)$$

Despite the fact that Eq. (12) has no general analytical solution [10], let us explore the asymptotic solution under the assumption *e* → 0, as follows:

$$\frac{d\Omega}{d\psi_p} = \left(-\frac{F(\psi_p)}{D}\right)\cdot\Omega^3 - \frac{E}{D}\cdot\Omega^2$$

$$D \cong 1,\ E \cong 0,\ F(\psi_p, f) \cong -\frac{\omega_0^2}{2}\cdot\sin\left(2(\psi_p + (p-1)\cdot f)\right) \Rightarrow$$

$$\frac{1}{\Omega^2} = \int 2F(\psi_p)\,d\psi_p \Rightarrow \Psi = \pm\sqrt{\left(\int 2F(\psi_p, f)\,d\psi_p\right)} = \pm\frac{\omega_0}{\sqrt{2}}\cdot\sqrt{C_0 + \cos\left(2(\psi_p + (p-1)\cdot f)\right)},$$

where $C_0 = const$, $C_0 > 1$. In addition, the previous equation yields the following:

$$\frac{d\psi_p}{df} = \Psi \Rightarrow \frac{d\psi_p}{df} = \pm\frac{\omega_0}{\sqrt{2}}\cdot\sqrt{C_0 + \cos\left(2(\psi_p + (p-1)\cdot f)\right)} \qquad (13)$$

Obviously, we can analytically resolve Eq. (13) only at $p = 1$:

$$\frac{d\psi_1}{df} = \pm\frac{\omega_0}{\sqrt{2}}\cdot\sqrt{C_0 + \cos 2\psi_1} \Rightarrow \int\frac{d\psi_1}{\sqrt{C_0 + \cos 2\psi_1}} = \pm\frac{\omega_0}{\sqrt{2}}\cdot\int df \qquad (14)$$

Nevertheless, we can see from Eqs. (13) and (14) that even with the appropriate simplifying assumptions $e \to 0$, $p = 1$, the asymptotical solution for the Beletskii's Eq. (1) seems not to be solved analytically (where $2\psi_1$ should be changed to $\delta$, but $\delta$ is the *doubled* angle between the radius vector of the center of mass and the appropriate axis of inertia of the satellite).

Meanwhile, we obtain the following from Eq. (14) ($p = 1$):

$$\cos 2\psi_p = u \implies \frac{1}{2}\int \frac{d(\arccos u)}{\sqrt{C_0 + u}} = \pm \frac{\omega_0}{\sqrt{2}} \cdot \int df \implies$$

$$-\frac{1}{2}\int \frac{du}{\sqrt{1-u^2} \cdot \sqrt{C_0 + u}} = \pm \frac{\omega_0}{\sqrt{2}} \cdot \int df \qquad (15)$$

The left part of Eq. (15) is known to be the proper *elliptical* integral in regard to the variable $u$ (see [15]). But the elliptical integral is known to be a generalization of a class of inverse periodic functions [16]. Thus, by obtaining the re-inverse dependence for Eq. (15), we could present the solution *as a set of quasi-periodic cycles*, that is, a quasi-periodic character of the evolution of the double angle $\psi_P$, which is the angle between the radius vector of the center of mass and the appropriate axis of inertia of the satellite.

## **Discussion**

Last but not least, we revisit the results of [6]. Equation (4) is known to be an *Abel* ODE of the second kind, a kind of generalization of the *Riccati* ODE. We also note that due to the special character of the solutions of *Riccati*-type ODEs [10], there is the potential for sudden *jumping* in the magnitude of the solution at some time parameter (or independent variable).

In the physical sense, this jumping of *Riccati*-type solutions of Eq. (4) could be associated with the effect of a sudden acceleration/deceleration of the satellite rotation around the chosen principle axis at a definite moment of parametric time. This means that there exists no chaotic regime of satellite rotation, but rather a potential for a kind of *gradient catastrophe* [17], depending on the initial conditions.

In general, we cannot exclude the possiblity that there is something in the aforementioned rotational dynamics of natural satellites. If such a chaotic phenomenon could occur, then it should be explored in future research to determine under which conditions it might occur, what this would imply for the rotational dynamics, and what assumptions this would require with respect to the orbit and shape of the satellite. We should also investigate whether it would be consistent with the assumption that obliquity can be neglected, and whether it is likely to actually happen in the solar system.

**Conclusion**

The main objective of this research was the analytical exploration of the dynamics of satellite rotation during its elliptical orbit motion around a planet.

To do so, we revisited the results of [6]. By elegantly changing the variables (considering the true anomaly *f* as the *independent* variable), we presented the governing equation of satellite rotation in the form of an *Abel* ODE of the second kind, a kind of generalization of the *Riccati* ODE.

In addition, we obtained the asymptotical solutions, which demonstrate a *quasi-periodic* character even with the strong simplifying assumptions $e \to 0$, $p = 1$, which reduces the governing equation of [6] to a kind of Beletskii's equation. We also suggest a revolving scheme for solving a cascade of *Abel* equations for satellite rotation, and test this scheme on the best candidates for calculations among the most massive satellites of the Uranus system (*see the proper calculations in the Appendix*).

**Acknowledgments**

**Appendix.**

**A.1.   Derivation of Eq. (3).**

We recall the relation between the true anomaly $f$ and the mean anomaly $\tau$ (in accordance with Kepler's law of orbital motion):

$$\frac{df}{d\tau} = \frac{(1+e\cdot\cos f)^2}{(1-e^2)^{\frac{3}{2}}} \qquad (16)$$

where $\tau = n\cdot t$, with $n$ being the mean motion of the satellite:

$$n^2 = \frac{G(M+m_0)}{a^3}, \qquad (17)$$

where the units have been chosen such that the orbital period of the satellite is $2\pi$ and its semi-major axis in Eq. (17) is equal to one (the mean motion is defined by the formula $n = 2\pi/P$, where $P$ is the period of the body motion in orbit). Thus, the dimensionless time $t$ is equal to the mean anomaly $\tau$.

We can derive from Eqs. (2) and (16) the following:

$$\frac{d\vartheta}{dt} = \frac{d\psi_p}{df}\cdot\frac{df}{dt} + p\cdot\frac{df}{dt}$$

$$\frac{d^2\vartheta}{dt^2} = \frac{d^2\psi_p}{df^2}\cdot\left(\frac{df}{dt}\right)^2 + \frac{d\psi_p}{df}\cdot\frac{d^2f}{dt^2} + p\cdot\frac{d^2f}{dt^2} \Rightarrow$$

$$\left(\frac{df}{dt} = \frac{(1+e\cdot\cos f)^2}{(1-e^2)^{\frac{3}{2}}},\quad \frac{d^2f}{dt^2} = \frac{2(1+e\cdot\cos f)\cdot(-e\cdot\sin f)}{(1-e^2)^{\frac{3}{2}}}\cdot\frac{df}{dt}\right)$$

$$\Rightarrow \frac{d^2\psi_p}{df^2}\cdot\left(\frac{(1+e\cdot\cos f)^2}{(1-e^2)^{\frac{3}{2}}}\right)^2 + \left(\frac{d\psi_p}{df}+p\right)\cdot\frac{2(1+e\cdot\cos f)\cdot(-e\cdot\sin f)}{(1-e^2)^{\frac{3}{2}}}\cdot\frac{(1+e\cdot\cos f)^2}{(1-e^2)^{\frac{3}{2}}} +$$

$$+ \frac{\omega_0^2}{2r^3}\sin(2(\psi_p+(p-1)\cdot f)) = 0 \qquad (18)$$

So, we can then properly transform Eq. (18), which should yield the following:

$$(1+e\cdot\cos f)\cdot\frac{d^2\psi_p}{df^2} - 2e\cdot\sin f\cdot\left(\frac{d\psi_p}{df}+p\right)+$$

$$+\frac{\omega_0^2}{2r^3}\left(\frac{1-e^2}{1+e\cdot\cos f}\right)^3 \sin(2(\psi_p+(p-1)\cdot f)) = 0 \qquad (19)$$

Compare Eq. (19) with Eq. (3) to observe the following difference:

$$(1+e\cos f)\cdot\frac{d^2\psi_p}{df^2} - 2e\sin f\cdot\left(p+\frac{d\psi_p}{df}\right)+$$

$$+\frac{\omega_0^2}{2}\cdot\sin\left(2(\psi_p+(p-1)\cdot f)\right) = 0,$$

and recall that in Eq. (6), as well, that the units have been chosen so that the expression below is equal to one ($a = 1$):

$$\frac{1}{r^3}\left(\frac{1-e^2}{1+e\cdot\cos f}\right)^3 = \frac{1}{r^3}\left(\frac{r}{a}\right)^3 = 1 \qquad (*)$$

### A.2. **Estimations in the revolving scheme for solving cascade of Abel equations**

Let us estimate the required interval $\Delta f$, for which the absolute meaning of the true anomaly $f$ could be considered to be *a constant* parameter when resolving Eq. (4).

For example, let us consider the case of the Titania satellite of Uranus (its most massive satellite). According to the formulae (10) and (11), using actual astrometric observation data for our calculations (see [18]), we obtain the following:

$$\Delta f \cong \frac{\alpha}{2e} = \frac{\alpha}{2 \cdot (0.00180148)},$$

$$\alpha = \frac{15}{4} \frac{GM}{Gm} \left(\frac{R}{a}\right)^3 \frac{k_2}{Q} = \frac{15}{4} \frac{5\,793\,939.3}{235.3} \left(\frac{788.9}{436\,281.94}\right)^3 \frac{0.03}{100}$$

$$\Rightarrow \quad \Delta f \cong \frac{\alpha}{2e} = \frac{\alpha}{2 \cdot (0.00180148)} = 4.55 \cdot 10^{-5} \cong 9.4'' \qquad (20)$$

where we choose $k_2 \cong 0.03$, $Q \cong 100$ [14]. The proper interval $\Delta f$ in Eq. (20), which is required for calculations when resolving a cascade of Abel Eqs. (4) for satellite rotation, corresponds roughly to the 125-km path on the elliptical orbit of Titania.

Let us also consider the cases of other massive satellites of Uranus:

1) The case of satellite Ariel:

$$\alpha = \frac{15}{4} \frac{GM}{Gm} \left(\frac{R}{a}\right)^3 \frac{k_2}{Q} = \frac{15}{4} \frac{5\,793\,939.3}{90.3} \left(\frac{577.9}{190\,929.79}\right)^3 \frac{0.03}{100}$$

$$\Rightarrow \quad \Delta f \cong \frac{\alpha}{2e} = \frac{\alpha}{2 \cdot (0.00136551)} = 73.3 \cdot 10^{-5} \cong 151.2'' \qquad (21)$$

which corresponds with the 879-km path on its quasi-circle orbit (orbital velocity is 5.5 km/s).

2) The case of satellite Umbriel:

$$\alpha = \frac{15}{4} \frac{GM}{Gm} \left(\frac{R}{a}\right)^3 \frac{k_2}{Q} = \frac{15}{4} \frac{5\,793\,939.3}{78.2} \left(\frac{584.7}{265\,984.01}\right)^3 \frac{0.03}{100}$$

$$\Rightarrow \quad \Delta f \cong \frac{\alpha}{2e} = \frac{\alpha}{2 \cdot (0.00424068)} = 10.44 \cdot 10^{-5} \cong 21.5'' \qquad (22)$$

which corresponds roughly with the 174-km path on its elliptical orbit.

3) The case of satellite Oberon:

$$\alpha = \frac{15}{4}\frac{GM}{Gm}\left(\frac{R}{a}\right)^3\frac{k_2}{Q} = \frac{15}{4}\frac{5\,793\,939.3}{201.1}\left(\frac{761.4}{583449.53}\right)^3\frac{0.03}{100}$$

$$\Rightarrow \quad \Delta f \cong \frac{\alpha}{2e} = \frac{\alpha}{2\cdot(0.00140798)} = 2.56\cdot10^{-5} \cong 5.3'' \qquad (23)$$

which corresponds roughly with the 94-km path on its quasi-circle orbit.

4) The case of satellite Miranda:

$$\alpha = \frac{15}{4}\frac{GM}{Gm}\left(\frac{R}{a}\right)^3\frac{k_2}{Q} = \frac{15}{4}\frac{5\,793\,939.3}{4.4}\left(\frac{234.2}{129848.11}\right)^3\frac{0.03}{100}$$

$$\Rightarrow \quad \Delta f \cong \frac{\alpha}{2e} = \frac{\alpha}{2\cdot(0.00132732)} = 327.4\cdot10^{-5} \cong 675.4'' \qquad (24)$$

which corresponds roughly with the 2671-km path on its quasi-circle orbit (orbital velocity is approximately 6.7 km/s).

As we can see, the best candidates for calculations (by the revolving scheme for solving a cascade of *Abel* equations for satellite rotation) are Ariel and Miranda: the time interval for the Ariel calculations is approximately (879 km/5.5 km/s) $\cong$ 160 s, and for Miranda appears to be approximately (2671 km/6.7 km/s) $\cong$ 400 s. This means 8,573 iteration steps in the calculations for Ariel (by the aforementioned scheme) per one rotation period and 1,919 iteration steps for Miranda.